\documentclass[twocolumn,showpacs,preprintnumbers,amsmath,amssymb]{revtex4}

\usepackage{graphicx}% Include figure files
\usepackage{dcolumn}% Align table columns on decimal point
\usepackage{bm}% bold math

\def\lapproxeq{\lower .7ex\hbox{$\;\stackrel{\textstyle
<}{\sim}\;$}}
\def\gapproxeq{\lower .7ex\hbox{$\;\stackrel{\textstyle
>}{\sim}\;$}}

\begin{document}

\preprint{}

\title{On the energy determination of extensive air showers\\ 
through the fluorescence technique}

\author{J. Alvarez-Mu\~niz}
\author{E. Marqu\'es}
\author{R.A. V\'azquez}
\author{E. Zas}
\affiliation{Departamento de F\'\i sica de Part\'\i culas, 
Facultade de F\'\i sica,\\
Universidade de Santiago de Compostela, 15706 Santiago, SPAIN\\
}

\begin{abstract}
The determination of the shower development in air using fluorescence
yield is subject to corrections due to the angular spread of the
particles in the shower. 
This could introduce systematic errors in the energy determination of
an extensive air shower through the fluorescence technique. 
\end{abstract}

% PACS meaning
% 96.40.Pq Extensive air showers
% 96.40.-z Cosmic rays


\maketitle

The fluorescence technique 
consists on observing the atmospheric nitrogen fluorescence
light induced by the passage of the charged particles in a shower through
the atmosphere. It is an alternative for ultra high energy cosmic ray 
(UHECR) detection to the more common experiments that detect the shower front 
as it reaches ground level. 
The technique was first explored in a pioneering experiment 
by the Fly's Eye detector \cite{Baltrusaitis88} and is currently
being exploited in its successor, the high resolution Fly's Eye 
(HiRes) \cite{Abu-Zayyad00}, as well as in the Auger 
Observatory \cite{Cronin95}
now in construction stage, and the Telescope Array \cite{TA}. 
It is also the basis of 
planned experiments that will look for atmospheric UHECR showers from 
satellites \cite{EUSO}, and which are likely to provide the next 
generation of UHECR detectors. 
Although the technique has been completely successful, there are unsolved 
discrepancies between the UHECR spectrum measured with it and with the 
more conventional 
techniques that sample the shower front as it hits the ground. 
The measurements of the cosmic ray spectrum above $5~10^{19}~$eV with the 
HiRes experiment give results which differ at the 2 to 3 sigma level from 
those obtained with the AGASA air shower array \cite{AGASA}. 
In this respect the Auger observatory, being a hybrid experiment that combines 
both the fluorescence and the shower front sampling techniques, is a crucial 
step for resolving the discrepancies and for the future of UHECR measurements. 

Fluorescence detectors use mirrors to collect light produced by all the 
charged particles in 
the shower front as it propagates through the atmosphere. Different depths 
of the shower are observed from different arrival directions which are
viewed by different photo-detectors of an imaging camera in the focal plane 
of the mirror. A procedure 
must be devised to compare the light emission curve as a function of depth 
to that expected from an air shower, from which important properties such 
as shower energy can be inferred. As fluorescence technique experiments 
measure the light emitted at different depths in shower development, 
they perform a calorimetric measurement of energy 
deposition, and are in principle more reliable for establishing shower energy. 

The technique has other difficulties, particularly those associated 
to light transport in the atmosphere which is complex and changing with time, 
and also because the fluorescence light emission mechanism itself is subject to 
uncertainties. In this article we discuss the relation between the 
emission process and the shower development. We argue that there are geometric 
effects associated to shower development that can have important implications  
in the procedure to compare light emission to shower development, with 
important consequences for the determination of shower energy. This idea 
is not new, it was already discussed over a decade ago 
by M.~Hillas \cite{hillas}.

The depth distribution of the collected photons 
in a fluorescence telescope 
$N^{\rm tot}_\gamma(X)$, where $X$ is measured along shower axis, 
is often converted to the number of charged particles $N_e(X)$ at the 
corresponding depth.  
In this way the longitudinal shower profile can be obtained after correcting 
for attenuation and geometrical effects associated to the characteristics
of the detector and the orientation of the shower with respect
to it,
\begin{equation}
N_e(X) =
\frac{N^{\rm tot}_\gamma (X)}{Y \Delta X}
\frac{1}{g_{\rm atten} g_{\rm area}},
\label{eq:size}
\end{equation} 
where $g_{\rm atten}$ takes into account the 
attenuation of the fluorescence photons in the atmosphere,
and $g_{\rm area}$ accounts for the collection area of the mirror  
(see Eq.(6.2) in Ref.\cite{Sokolskybook}). Here $\Delta X$ is the 
segment of the shower viewed by the telescope which is measured along 
the shower axis, and $Y$ is the fluorescence yield for air.   

The amount of fluorescence light emitted by a particle of charge $e$ 
is usually assumed to be isotropic and proportional to the energy 
loss by ionization.
%, which is in turn known to be proportional to the pathlength it travels. 
The fluorescence yield for nitrogen, $Y$, measured in photons/m, 
is concentrated in the Ultra Violet (UV) range of the spectrum, and must be 
experimentally determined.  
Most of the data used relies on two experiments \cite{tesis,Kakimoto95}. 
In the most recent experiment, 
electrons were fired onto an air target of a given length. 
The yield is obtained dividing the number of emitted photons per incident 
electron ($N_\gamma$) by the length ($d$) of the visible portion of the 
electron beam in the direction of the beam axis \cite{Kakimoto95}
\begin{equation}
Y_{\rm exp}=\frac{N_\gamma}{d}.
\label{eq:yield}
\end{equation}
The experimental result is that on average about five UV fluorescence 
photons are emitted when a particle of charge $e$ passes through one meter
of air at standard temperature and pressure \cite{Kakimoto95}. 

The shower energy has been determined in the past by integration of the 
longitudinal profile of the shower \cite{Sokolskybook,Song00},   
applying Eq.~\ref{eq:size} to obtain it, 
and fitting it  
to an adequate depth development function \cite{Abu-Zayyad01}. 
The numerical value of this integral is just the  
track length of the charged particles in the shower. Since the integral 
is performed in the direction of the shower axis, the calculated track length 
corresponds to the sum of all the charged particle track lengths projected
onto the shower axis. We have referred to this track length as the 
{\it total projected track length} in a different context \cite{ZHS91}. 
This track length is known to be proportional to the shower energy,
\begin{equation}
E_{\rm em}=\alpha \int_0^{\infty} N_e(X)dX,
\label{eq:Eem}
\end{equation}
where $\alpha$ is a constant that can be related to the  
ionization loss rate. Its numerical value has been obtained 
from Monte Carlo
simulations \cite{Song00} and is usually given the value 
$\alpha\sim 2.19~{\rm MeV/g~cm^{-2}}$. 
%close to the minimum ionization energy loss.
We have performed a detailed simulation of 1 TeV electromagnetic
showers in air using the GEANT 4 \cite{GEANT4} package, following particles
down to a very low kinetic energy threshold of 10 keV. 
These simulations produced a numerical value $\alpha=2.25~{\rm MeV/g~cm^{-2}}$.
For the sake of clarity we adopt in the following the standard
value of $\alpha=2.19~{\rm MeV/g~cm^{-2}}$, keeping in mind 
that any uncertainty in $\alpha$ translates directly into a systematic error
in the determination of shower energy. It is also important to note
that an accurate determination of $\alpha$ by Monte Carlo simulations
should take care for the backward-going particles in the shower. We 
have obtained that $\sim 3-4\%$ of the contribution to the track integral
in Eq.~\ref{eq:Eem} is due to backscattered particles. 

The main idea of this article refers to the relation between energy loss in 
a shower and the value of the projected track length. It is well known 
that as the shower develops charged particles degrade their energy, they 
suffer more scatterings and they acquire transverse momentum. In other 
words their paths become less aligned with shower axis. However both the 
energy loss and the fluorescence yield of charged particles are defined 
per unit length measured along the particle travel 
direction. If the fluorescence light is used to infer the number of shower 
particles in a given depth interval, and the fluorescence yield is assumed 
to be proportional to energy loss, account must be taken that the energy loss 
of particles depends on $\Delta L$, the average track length traveled by the 
shower particles {\it regardless of the particle direction} in a given depth 
interval $\Delta X$ measured along the shower axis. 
While $\Delta L$ is measured along the particle direction, 
$\Delta X$ is measured along the shower axis. 
One expects the ratio of $\Delta L$ to $\Delta X$ to increase as the shower 
develops and the average energy of the shower particles decreases. 
This ratio can be thought as the average value of $\sec\theta$ for 
all tracks, $\theta$ being the angle between the particle direction
and the shower axis.

\begin{figure}
\centerline{\includegraphics[width=8.5cm]{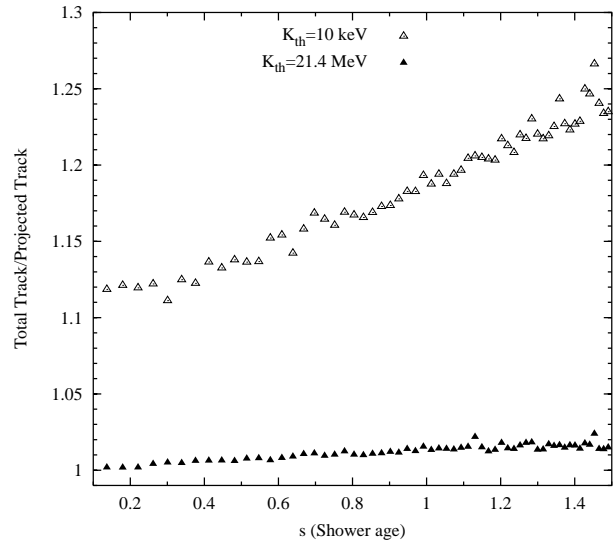}}
\caption{Ratio of total and projected track lengths ($f$) in electromagnetic
showers as a function of depth in the atmosphere (expressed as shower
age $s=3t/(t+2\beta)$ where $\beta=\log(E/E_c)$ and $E_c$ is the critical
energy in air). Empty (filled) symbols are calculated for kinetic 
threshold energy of
10 keV (21 MeV) and shower energy 1 TeV.}
\label{fig:fig1}
\end{figure}

We have performed shower simulations in air to calculate the relation 
between $\Delta L$ and $\Delta X$ as a function of atmospheric depth. 
Air showers initiated 
by photons have been simulated using the GEANT 4 \cite{GEANT4} and the
ZHS \cite{ZHS91} packages. It is worth noting that we have obtained 
that the results of both 
simulations in air, under the same conditions, agree at the $1\%$ level.  
We define a factor $f$ as the ratio of the total to
the projected track length. 
We have found that $f$ in a shower is significantly different from 1. 
Moreover it increases with shower age as expected. 
In Fig~\ref{fig:fig1} we plot the dependence on depth 
(expressed in terms of shower age) of $f$ as obtained  
in electromagnetic showers for 1 TeV showers. 
%
%%%%%%%%%%%%%%%%%%%%%%%%%%%%%%%%%%%%%%%%%%%%%%%%
%%%%%%%%%%%%%%%%%%%%%%%%%%%%%%%%%%%%%%%%%%%%%%%%
The effect can amount to a $25\%$ ($f=1.25$) correction at the end of 
the shower. We have also obtained that the correction is energy 
independent which is not surprising since showers 
are known to have very good scaling properties. 
A reference value of the effect is given by $f=1.18$ at shower
maximum $s=1$. 

\begin{figure}
\centerline{\includegraphics[width=8.5cm]{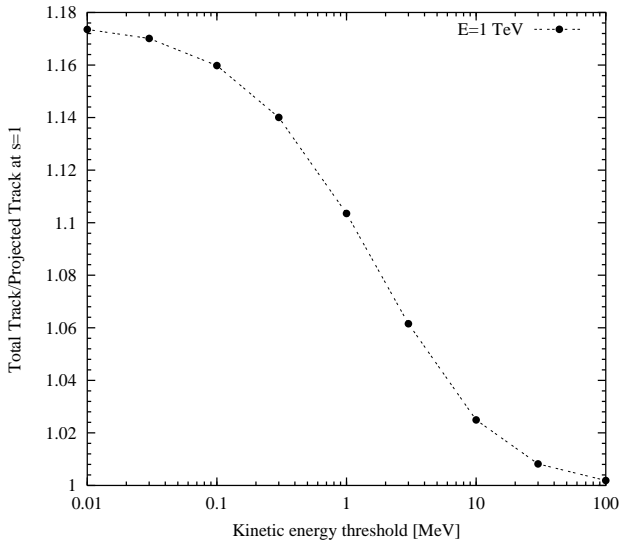}}
\caption{Ratio of total and projected track lengths ($f$) in electromagnetic
showers at maximum as a function of the threshold energy.}
\label{fig:fig2}
\end{figure}

It is worth noting that this effect is not seen in experiments that measure
the \v Cerenkov light emission from a shower. The kinetic threshold energy for 
\v Cerenkov emission in air is around 21 MeV. At this energy the deflection 
angle is very small and the correction (expressed as $(f-1)\times 100$) 
is less than 1\%.
We also show in Fig.~\ref{fig:fig1} the calculated value of $f$ for
showers with threshold energy of 21 MeV. The ratio is seen to be 
very small at all stages of shower development. 
In figure \ref{fig:fig2} we show $f$  
at shower maximum as a function of the kinetic energy threshold 
introduced in the simulation. At a kinetic  
energy threshold of 3 MeV the correction is around 1.06 and goes
rapidly to 1 at higher energies. For a kinetic energy threshold of 10~keV 
it increases to about 1.18. 
%%%%%%%%%%%%%%%%%%%%%%%%%%%%%%%%%%%%%%%%%%%%%%%%
%%%%%%%%%%%%%%%%%%%%%%%%%%%%%%%%%%%%%%%%%%%%%%%%
For showers initiated by hadrons the effect must be similar because they can 
be considered as a superposition of electromagnetic showers initiated from 
photons from neutral pion decays at different depths. To a reasonably good
approximation the lateral distribution of hadronic showers around maximum 
does not change very much with depth and corresponds to that of an
electromagnetic shower of age $s\simeq 1$ \cite{alvarezPLB98}. 
We can expect $f\simeq 1.18$ for these showers around shower maximum. 

%The effect discussed here can be also addressed by considering an alternative 
%to Eq.~\ref{Eem} in which the right hand side is the total track length of the 
%shower. 
%%
%\begin{equation}
%E_{\rm em}=\bar \alpha \sum {\rm track lengths},
%\label{eq:Etk}
%\end{equation}
%%  
%It is relatively easy to show by simulation that in that case the 
%proportionality constant is reduced accordingly to 
%$\bar \alpha \sim 1.7~{\rm MeV/g~cm^{-2}}$ instead of the widely used value 
%of $2.19~{\rm MeV/g~cm^{-2}}$. 
%This is the approach taken in Ref.~\cite{Hillas} although the numerical 
%values quoted for these constants are different. 
%This effect was more recently noticed in Ref.~\cite{Risse} where the ratio 
%of energy deposit and the number of charged particles in the showers computed
%using full Monte Carlo simulation is shown to be $\sim 15\%$ higher than
%expected. However no explanation of the discrepancy was given. 

The same effect has to be kept in mind in experiments that 
measure the photon yield applying Eq.~\ref{eq:yield}. 
The electrons traveling through Nitrogen a distance $d$ will be 
scattered, and as a result the path they travel will be longer 
than its projection onto the beam axis depending on the electron's
energy. 
Providing that $d$ is sufficiently short, an electron does not deviate much
from the incident direction, and its total track length is approximately 
equal to the length of the visible portion of the beam. 
The ratio of the track lengths will actually depend on electron energy and 
must also be taken into account in the calculation of the yield. 
We have simulated electron paths 
in air using the GEANT 4 package, propagating a 1.4 MeV electron 
along a distance of 30 cm  along the direction of movement of the 
incident electron, to roughly reproduce the conditions of the experimental 
setup in \cite{Kakimoto95}. 
These simulations predict an average ratio $f_{\rm exp}=1.02$
between the total track length and the track length projected along the 
electron's incident direction. 
This value is significantly smaller than the typical value of $f$ obtained in 
shower simulations and as a result corrections should be made to account for 
these factors. 

Assuming that the fluorescence light emitted is just proportional to the 
track length of charged particles in a shower, 
we introduce an effective value of the yield 
($Y_{\rm eff}(X)$) which is dependent on depth and takes into consideration 
the fact that charged particles do not travel parallel to shower axis. 
Let us consider that the fluorescence yield $Y_{\rm exp}$ is obtained 
in a given 
experimental setup for which the ratio of total to projected track 
lengths is given by $f_{\rm exp}$. The effective yield 
that should be used in air shower experiments is given by:
\begin{equation} 
Y_{\rm eff}(X)= {f(X) \over f_{\rm exp}} ~Y_{\rm exp}. 
\label{eq:yieldeff}
\end{equation}
In other words $Y_{\rm exp}$ must be corrected by a factor $f(X)/f_{\rm exp}$
which is $\sim 1.16$ near maximum, assuming $f_{\rm exp}=1.02$. 

The deduced value of the number of shower particles, $N_e(X)$, must be 
inversely proportional to the effective yield which now measures the emission 
in terms of depth along the shower axis. 
Since $f(X)>f_{\rm exp}$ for most depths, using the experimental yield without
this correction to estimate the shower energy will lead to an overestimate 
of the electromagnetic shower energy.  
This translates directly into a systematic overestimate of the energy 
which is deposited in the atmosphere in the form of charged 
particles (mainly electrons and positrons). 

Moreover since $Y_{\rm eff}$ is 
in general depth dependent, ignoring the correction factor can also affect 
other observables such as the position of shower maximum 
We can roughly estimate an upper bound of the expected effect on the 
calculation of the depth of shower maximum. 
The number of particles as a function of depth obtained without correcting 
for this effect would be 
$\tilde N_e(X) = f(X) N_e(X)$, where $N_e(X)$ is the actual number 
of particles. 
Using a simple parameterization of $f$ from figure 
\ref{fig:fig1} and the Greisen parameterization for $N_e(X)$ one
obtains a depth of maximum which is deeper by 
$\Delta X_{\rm max} \sim 4 $ g/cm$^2$.
This effect was already noticed in \cite{Risse}. Our upper bound 
estimate is compatible with the shift of $3-4 ~{\rm g/cm^2}$ between 
the depth at which the fluorescence light emission reaches maximum
and the depth at which $N_e$ is maximum, obtained in \cite{Risse} 
for hadronic showers. 

The result is independent of energy. This implies, for instance, that the
elongation rate, the derivative of $X_{\rm max}$ with respect to
the logarithm of the energy, is unchanged by this effect. 
Finally, the change of $f$ with depth for showers induced by hadrons is 
expected to be less important because the 
lateral distribution function of hadronic showers is known to be less 
dependent on depth. In any case an iterative process taking this small effects 
into account could be implemented to avoid such systematic effects. 

We have shown that the fact that charged particles in a shower do not travel 
parallel to the shower axis has important implications in the fluorescence 
light output of a shower in the assumption that the fluorescence yield is only
dependent on distance traveled by the individual particles. 
The light output per unit charged particle (as measured in the plane 
containing the shower axis perpendicular to the direction of observation) 
increases as the shower develops in the 
atmosphere since the
individual particles travel in larger angles with respect to the shower axis
as the shower develops. 
In a conventional approach the deduction of the depth development curve of the 
shower from the detected light needs to take these effects into account. 
This can be made trough an effective yield which is to be obtained in a 
detailed simulation. 
Proposals to directly compare the data to the light outputs generated by 
simulated showers are at an advantage to take such effects into
consideration in a consistent way \cite{GuerardGAP}. 
Alternatively it has been recently proposed to directly measure the energy 
deposited in fluorescence light in order to estimate the shower energy 
\cite{Dawson}. Such an energy determination would also not be subject to 
the corrections discussed here. However, any determination of the 
number of particles through the fluorescence technique
has to be corrected by the factor discussed here.

{\bf Acknowledgments} 
We thank P. Privitera, P. Sommers, and A.A. Watson for
helpful insights and suggestions after carefully reading
the manuscript. We especially thank P. Sommers
for pointing out the potential problem with the \v Cerenkov 
light. This work was partly supported by 
the Xunta de Galicia (PGIDIT02 PXIC 20611PN), by MCYT 
(FPA 2001-3837 and FPA 2002-01161).
R.A.V. is supported by the ``Ram\'on y Cajal'' program.
We thank CESGA, ``Centro de Supercomputaci\'on de Galicia'' for
computer resources.
%%%%%%%%%%%%%%%%%%%%%%%%%%%%%%%%%%%%%%%%%%%%%%%% 

\end{document}